\documentstyle[aps,prd]{revtex}

\tighten

\def\D {\mbox{D}}
\def \t {\tilde}
\def\div {\mbox{div}\,}
\def\rd {\displaystyle{\cdot}}
\def\c {\mbox{curl}\,}
\def\ep {\varepsilon}
\def \ts {\textstyle}
\def\be {\begin{equation}}
\def\ee {\end{equation}}
\def\bea {\begin{eqnarray}}
\def\eear {\end{eqnarray}}
\def\la {\langle}
\def\ra {\rangle}

\begin{document}

\title{Covariant velocity and density perturbations in
quasi-Newtonian cosmologies}

\author{Roy Maartens}

\address{School of Computer Science and Mathematics,
Portsmouth University, Portsmouth~PO1~2EG, Britain}

\date{August 1998}

\maketitle

\begin{abstract}

Recently a covariant approach to cold matter universes
in the zero-shear hypersurfaces (or longitudinal)
gauge has been developed. This
approach reveals the existence of 
an integrability condition, which does not appear in
standard non-covariant treatments.
A simple derivation and generalization of the
integrability condition is given, 
based on showing that the quasi-Newtonian models are a
sub-class of the linearized `silent' models. The
solution of the integrability condition implies a propagation 
equation for the acceleration. It is shown how
the velocity and density perturbations
are then obtained via this propagation equation.
The density perturbations acquire a small relative-velocity
correction on all scales, arising from
the fully covariant general relativistic analysis.

\end{abstract}

\pacs{04.20.Nx, 98.80.Hw, 98.65.Dx, 95.30.Sf}

\section{Introduction}

Density and velocity perturbations of cold matter universes
are crucial to the understanding of structure formation in
cosmology \cite{p}. On scales well below the Hubble length,
Newtonian theory may be used to analyze gravitational
instability. However,
current and upcoming observations and simulations are
probing scales which are a significant fraction of the Hubble
length, thus requiring a relativistic treatment \cite{tf}.
Bardeen's gauge-invariant theory \cite{b} is most often used
for relativistic perturbations. Various choices of gauge are
possible in the theory. 
The zero-shear hypersufaces (or longitudinal)
gauge \cite{b,b2} sets up
a frame of reference which emulates that of Eulerian observers
in Newtonian theory, thus motivating the term `quasi-Newtonian'.
Since the frame is non-comoving, there are relative-velocity
effects on the density perturbations, and subtle 
issues arise in dealing with these effects, as pointed out
recently by van Elst and Ellis \cite{vee}. In order to resolve
these problems in a fully
gauge-invariant way, a covariant approach may be adopted.

The covariant and gauge-invariant
approach to perturbations was developed by Ellis and Bruni \cite{eb} 
on the basis of Hawking's paper \cite{h}. It has
various advantages over the non-covariant gauge-invariant theory
(see \cite{bde} for further discussion). 
One advantage is that all quantities have a direct and immediate
physical or geometric meaning, and no nonlocal decomposition
into scalar, vector or tensor modes is required.
A second advantage is that the covariant approach provides a natural
and transparent setting to search for integrability conditions 
which may arise from constraints.
These general relativistic constraints and the consequences
that follow from their evolution, are often not made
explicit. As a result, crucial general relativistic effects
can sometimes be obscured or missed.

Covariant consistency analysis of constraints was
developed by Maartens \cite{m}, building on the
methods of Lesame et al. \cite{lde}, in a form applicable to both the
nonlinear exact theory and the case of linearized perturbations. 
(Further developments of the approach are given
in \cite{ve,v,mac}.) Applications of a covariant approach to
`silent' universes \cite{mn,veplus,mle2},
to nonlinear gravitational radiation \cite{mle,smel}, and
to non-accelerating fluid models \cite{sss,s}
reveal the existence of crucial integrability conditions.
The covariant characterization of 
scalar, vector and tensor perturbations
also relies on such an approach \cite{dbe,he,mes,tb,cl}.

Van Elst and Ellis \cite{vee}
use a covariant consistency analysis first
in nonlinear quasi-Newtonian cosmologies and then in
the case of covariant
linearization about a Friedmann-Lemaitre-Robertson-Walker
(FLRW) background. They show that the nonlinear models are
likely in general to be inconsistent, 
except for special
cases such as FLRW solutions. 
Inconsistency of the nonlinear models is not surprising, since
one 
is demanding for all possible dynamical evolutions of matter
that there exists a shearfree and irrotational congruence,
which in particular forces the magnetic part of the Weyl curvature
to vanish \cite{t}. This rules out gravitational 
radiation \cite{veplus}, and thus
leads to severe restrictions on
the gravitational field \cite{veplus,s2,kp}.
In the linearized case, one might expect that all the problematic
terms which arise in evolving the constraints would be removed.
It is implicitly or effectively assumed in standard non-covariant
perturbation theory that there are no integrability
conditions arising from the zero-shear hypersurfaces gauge.

However, it turns out that the linearized models are not
in general consistent.
Van Elst and Ellis introduce
an ansatz for the evolution of the gravitational
potential, which they motivate by a discussion of the lapse 
in ADM-type approaches.
Using this ansatz, they find
an integrability condition in the linearized models. The
integrability condition
is satisfied by a particular value 
of the constant parameter in the ansatz.\footnote{
For other values of the parameter, only very special
models appear to be consistent \cite{vee}.}
The reason that the integrability condition is not revealed
in some non-covariant treatments is probably either
an implicit assumption
that only evolution equations need be considered, 
or a gauge-dependent approximation that 
effectively neglects the velocity and 
removes the constraints (see \cite{vee} for further
discussion of this point).

In this paper, extensions of the results of \cite{vee}
on linearized quasi-Newtonian models
are given. The main result is the determination of the 
velocity and density perturbations. A crucial part of the analysis
is the combined use of the comoving (`Lagrangian')
and quasi-Newtonian (`Eulerian') frames.

Section II summarizes the necessary covariant
equations and methods, with some details given in appendices.
Section III uses a transformation to the comoving frame
in order to show that the quasi-Newtonian models are 
a sub-class of the linearized silent models. This is the basis
for a simple and direct approach to deriving the
integrability condition in general, i.e. without introducing
any ansatz for the gravitational potential. The general 
integrability condition reduces to the special form given 
in \cite{vee} when one imposes their ansatz for
the gravitational potential.
Their ansatz is also generalized.
Furthermore, another integrability condition is derived
by considering spatial consistency of the constraints.
The van Elst-Ellis solution 
of the first integrability condition is shown to 
reduce the second condition to an identity. 

The main result however, follows from the fact
that the van Elst-Ellis 
solution itself implies a crucial propagation
equation for the 4-acceleration. This propagation
equation then leads to an equation 
determining the velocity perturbations. 
The equation is scale-independent, so that velocity
perturbations have an effect on all scales.
The velocity perturbation equation is readily solved analytically
for a flat background.

In section IV, the density perturbations are found
by a direct and simple approach. The complicated
calculations in \cite{vee} of the energy flux (momentum
density) source term in the
density perturbation equation are by-passed by considering
the density perturbations in the comoving frame. 
Furthermore, the equation is solved for a flat background.
The covariant correction to density perturbations
that arises from relative-velocity effects is 
a simple comoving divergence term, which can 
be found from the velocity perturbations. This correction
affects all scales, although it is rapidly dominated by the
usual solution.
The correction to the growing mode is consistent with the result of
Takada and Futamase \cite{tf}, who use non-covariant theory, but not
in the zero-shear hypersurfaces gauge.

Concluding remarks are made in section V.

\section{Covariant equations}

Given a choice of 4-velocity field $u^a$ (with $u^au_a=-1$), the
Ehlers-Ellis approach \cite{eh,e} employs only fully covariant
quantities and equations with transparent physical and geometric
meaning. The quantities are split into
spacetime scalars and spatially projected tensors, while the equations
split into evolution equations along $u^a$ and constraint equations
involving only spatial covariant derivatives. These equations arise
from the Ricci identity for $u^a$ and the Bianchi identities,
with Einstein's field equations incorporated via algebraic
replacement of the Einstein tensor by the energy-momentum tensor
$T_{ab}$.
The covariant linearization of the equations is the basis for
the Ellis-Bruni perturbation theory \cite{eb}.
Integrability conditions, at both the nonlinear and linear levels,
arise from investigating the derivatives of constraint equations
(including additional conditions that may be imposed by physical
or geometric assumptions), using
covariant differential identities and the evolution equations.
The streamlined and developed version of the Ehlers-Ellis
formalism given by Maartens \cite{m} (see also \cite{mes,mb,mt})
greatly facilitates such investigations, especially by
making explicit the irreducible quantities and derivatives, 
which significantly simplifies the equations, and
by developing the covariant identities which these quantities 
and derivatives obey.

The projection tensor $h_{ab}=g_{ab}+u_au_b$, where $g_{ab}$ is the
spacetime metric, and the projected alternating tensor
$\ep_{abc}=\eta_{abcd}u^d$, where $\eta_{abcd}=-\sqrt{|g|}
\delta^0{}_{[a}\delta^1{}_b\delta^2{}_c\delta^3{}_{d]}$ is
the spacetime alternating tensor, are the basis for covariant 
irreducible splitting of tensors and derivatives.
Projected rank-2 tensors $S_{ab}$
are split into a scalar trace, a projected 
vector spatially dual to the skew part, and a projected symmetric 
tracefree part:
\[
S_{ab}={\ts{1\over3}}S_{cd}h^{cd}h_{ab}+\ep_{abc}S^c+S_{\la ab\ra}\,,
\]
where $S_a={\ts{1\over2}}\ep_{abc}S^{[bc]}=S_{\la a\ra}\equiv
h_{ab}S^b$ and
$S_{\la ab\ra}\equiv [h_{(a}{}^ch_{b)}{}^d-{\ts{1\over3}}h^{cd}
h_{ab}]S_{cd}$.
Covariant time and spatial derivatives are defined by
\[
\dot{S}^{a\cdots}{}{}_{b\cdots}=u^c\nabla_c
S^{a\cdots}{}{}_{b\cdots}\,,~
\D_cS^{a\cdots}{}{}_{b\cdots}=h_c{}^fh^a{}_d\cdots h_b{}^e\cdots 
\nabla_fS^{d\cdots}{}{}_{e\cdots}\,,
\]
and then 
the covariant spatial divergence and curl are \cite{m,mes}
\bea
&& \div V=\D^aV_a\,,~~~~~~(\div S)_a=\D^bS_{ab}\,, \\
&& \c V_a=\ep_{abc}\D^bV^c\,,~~ \c S_{ab}=\ep_{cd(a}\D^cS_{b)}{}^d \,.
\eear
The dynamic quantities are the energy density $\rho$, the 
pressure $p$, the 
energy flux $q_a=q_{\la a\ra}$, and the anisotropic 
stress $\pi_{ab}=\pi_{\la ab\ra}$, so that
\[
T_{ab}=\rho u_au_b+ph_{ab}+2q_{(a}u_{b)}+\pi_{ab} \,.
\]
The kinematic quantities are given by
\[
\nabla_bu_a = \D_bu_a-A_au_b \,,~~
\D_bu_a = {\ts{1\over3}}\Theta h_{ab}+\sigma_{ab}+\ep_{abc}
\omega^c\,,
\]
where 
$\Theta=\D^au_a$ is the expansion,
$A_a=\dot{u}_{a}=A_{\la a\ra}$ is the 4-acceleration,
$\omega_a=-{\ts{1\over2}}\c u_a=\omega_{\la a\ra}$ 
is the vorticity, and
$\sigma_{ab}=\D_{\la a}u_{b\ra }$ is the shear.
Finally, the
gravito-electromagnetic fields are
\[
E_{ab}=C_{acbd}u^cu^d=E_{\la ab\ra }\,,~~~
H_{ab}={\ts{1\over2}}\ep_{acd}C^{cd}{}{}_{be}u^e=H_{\la ab\ra}\,,
\]
where $C_{abcd}$ is the Weyl tensor, which represents the locally
free gravitational field \cite{e,mes}. The
FLRW background is then covariantly and gauge-invariantly 
characterized by:
dynamics -- $\D_a\rho=0=\D_a p$,
$q_a=0$, $\pi_{ab}=0$; kinematics -- $\D_a\Theta=0$, $A_a=0=\omega_a$,
$\sigma_{ab}=0$;
gravito-electromagnetic field -- $E_{ab}=0=H_{ab}$.

In this paper only the linearized quasi-Newtonian cosmologies are
considered, since the main focus here is on perturbations and
structure formation. The covariant linearized
evolution equations in the general case are \cite{mt}
\begin{eqnarray}
 \dot{\rho} &=& -(\rho+p)\Theta-\div q \,,
 \label{c3}\\
\dot{\Theta} &=& -{\ts{1\over3}}\Theta^2
-{\ts{1\over2}}(\rho+3p)+\div A \,,
 \label{a1}\\
 \dot{q}_{a} &=&
-4H q_a
-(\rho+p)A_a -\D_a p
-(\div\pi)_{a} \,,
\label{c4} \\ 
 \dot{\omega}_{a} &=& -2H\omega_a
-{\ts{1\over2}}\c A_a \,, \label{a2}\\
 \dot{\sigma}_{ab} &=& -2H\sigma_{ab}
-E_{ab}+{\ts{1\over2}}\pi_{ab} +\D_{\la a}A_{b\ra }\,,
\label{a3}\\
 \dot{E}_{ab} &=& -3H E_{ab}
+\c H_{ab}
-{\ts{1\over2}}
\dot{\pi}_{ab} 
-{\ts{1\over2}}(\rho+p)\sigma_{ab} 
-{\ts{1\over2}}\D_{\la a}q_{b\ra }-{\ts{1\over2}}H
\pi_{ab}                                              \,,
\label{a4}\\
\dot{H}_{ab} &=& -3H H_{ab}
-\c E_{ab}
+{\ts{1\over2}}\c\pi_{ab}  \,,
\label{a5}
\end{eqnarray}
and the linearized constraint equations are 
\begin{eqnarray}
{\cal C}^1{} &\equiv& \div\omega =0 \,,\label{a7}\\
{\cal C}^2{}_a &\equiv &
(\div\sigma)_{a}-\c\omega_a 
-{\ts{2\over3}}\D_a\Theta
+q_a =0  \,,\label{a6}\\
 {\cal C}^3{}_{ab} &\equiv &
 \c\sigma_{ab}+\D_{\la a}\omega_{b\ra }
 -H_{ab} =0 \,, \label{a8}\\
 {\cal C}^4{}_a &\equiv &
 (\div E)_{a}
+{\ts{1\over2}}(\div\pi)_{a}
 -{\ts{1\over3}}\D_a\rho
+H q_a
 =0 \,, \label{a9}\\
 {\cal C}^5{}_a &\equiv &
 (\div H)_{a}
+{\ts{1\over2}}\c q_a 
 -(\rho+p)\omega_a
 =0 \,,\label{a10}
\end{eqnarray}
where $H=\dot{a}/a$ 
is the background Hubble rate, related to the background
values of $\rho$ and $p$ by
\be
\rho=3\left(H^2+{{\cal K}\over a^2}\right)\,,~~
\dot{H}+H^2=-{\ts{1\over6}}(\rho+3p)\,,
\label{frw}\ee
with ${\cal K}=0,\pm 1$ the curvature index.
The differential identities that are needed for investigating
consistency and deriving perturbation equations 
are collected in Appendix A. 

If another 4-velocity $\t{u}^a$ is chosen, 
the corresponding
kinematic, dynamic and gravito-electromagnetic quantities
undergo transformations. For completeness, and since they do not
appear elsewhere, the exact nonlinear
form of these transformations is given in Appendix B. Only the 
linearized form of the expressions is required below.

The covariant
characterization of quasi-Newtonian cosmologies 
is as follows \cite{vee}:
they are almost-FLRW
dust universes, with a congruence of observers whose
4-velocity field $u^a$ is irrotational, shearfree and 
nonrelativistic relative to comoving observers. 
The comoving 4-velocity $\t{u}^a$
is given by the linearized form of Eq. (\ref{b1}):
\be
\t{u}^a=u^a+v^a \,,
\label{tuu}\ee
where $v^a$ is the nonrelativistic relative velocity, which
vanishes in the background.
The models are thus defined in the quasi-Newtonian frame by
\bea
&&\mbox{(dynamics) }~~  p=0\,,~q_a=\rho v_a\,,~\pi_{ab}=0\,, 
\label{1}\\
&&\mbox{(kinematics) } ~~ \omega_a=0\,,~\sigma_{ab}=0 \,,
\label{2}
\eear
as shown by the linearized form of equations (\ref{b3})--(\ref{b10}).
Thus quasi-Newtonian cosmologies 
are irrotational, shearfree dust spacetimes with 
energy flux (momentum density)
arising purely from a particle flux in the  
quasi-Newtonian frame relative to 
the comoving frame. Note that 
the isotropic and anisotropic stresses that arise from
relative motion are second order in $v_a$, 
as given by equations (\ref{b8}) and (\ref{b10}).

The gravito-magnetic constraint equation (\ref{a8}), together with
Eq. (\ref{2}), shows that
\be
H_{ab}=0 \,.
\label{2'}\ee
Thus there is no gravitational radiation \cite{veplus}, which
further justifies the term `quasi-Newtonian'.
In addition, the div $H$ constraint (\ref{a10}), together with
Eq. (\ref{2}), shows that $q_a$ is irrotational, and thus
so is $v_a$:
\be
\c v_a=0=\c q_a \,.
\label{2''}\ee
Since the vorticity vanishes, it follows (see \cite{b2}) that
$v_a=\D_a\psi$, where
the velocity potential $\psi$ is determined below.

\section{Integrability conditions}

In irrotational dust models with vanishing energy flux
and anisotropic stress, the constraint equations 
${\cal C}^A=0$ evolve consistently with the evolution equations, 
even at the nonlinear level, in the sense that
\cite{m} (see also \cite{ve,v,mac})
\[
\dot{\cal C}^A=F^A{}_B\,{\cal C}^B+G^A{}_{Ba}\,\D^a{\cal C}^B\,,
\]
where $F$ and $G$ depend only on the kinematic, dynamic 
and gravito-electromagnetic quantities (and not their
derivatives).
If one imposes the `silent' constraint (\ref{2'}), then the nonlinear
models are generically inconsistent, but the linearized models
are consistent \cite{mn,veplus}. A very simple approach to the
integrability conditions for quasi-Newtonian cosmologies follows from
showing that these models are in fact a sub-class of the linearized
silent models. This can be seen by transforming to the comoving frame.

Linearizing the expressions in appendix B for the 
case where $u^a$ and $\t{u}^a$ are any frames in
nonrelativistic relative motion, one
finds for the kinematic quantities
\bea
\t{\Theta} &=& \Theta+\div v \,,\label{7}\\
\t{A}_a &=& A_a+\dot{v}_a+H v_a \,,\label{8} \\
\t{\omega}_a &=& \omega_a-{\ts{1\over2}}\c v_a \,,\label{9}\\
\t{\sigma}_{ab} &=& \sigma_{ab}+\D_{\la a}v_{b\ra} \,,\label{10}
\eear
for the dynamic quantities
\be
\t{\rho}=\rho\,,~~\t{p}=p\,,~~\t{q}_a=q_a-(\rho+p) v_a\,,~~
\t{\pi}_{ab}=\pi_{ab}\,,
\label{13}\ee
and for the gravito-electromagnetic field
\be
\t{E}_{ab}=E_{ab}\,,~~\t{H}_{ab}=H_{ab}\,.
\label{14}\ee
For $u^a$ the quasi-Newtonian frame and $\t{u}^a$
the comoving frame,
it follows from the equations (\ref{1})--(\ref{14}) that
\bea
&&\t{p}=0\,,~~\t{q}_a=0\,,~~\t{\pi}_{ab}=0 \,,\label{q5}\\
&&\t{A}_a=0\,,~~
\t{\omega}_a=0\,,~~
\t{\sigma}_{ab}=\D_{\la a}v_{b\ra}\,, \label{12}\\
&&\t{E}_{ab}=E_{ab}\,,~~
\t{H}_{ab}=0\,. \label{q6}
\eear
Equations (\ref{q5})--(\ref{q6}) constitute a covariant
characterization of linearized silent universes, except that
the shear takes a special form. Thus quasi-Newtonian models
are linearized silent models with a special form of shear, and
integrability conditions arise only from the restriction on the 
shear.

It is now convenient to return to the quasi-Newtonian frame,
where the restriction on the shear is that it vanishes.
Integrability conditions arise directly from the fact that the 
shear propagation equation
(\ref{a3}) is turned into a constraint, i.e. $E_{ab}=\D_{\la a}
A_{b\ra}$. This can be simplified, using
$\c A_a=0$, which follows from the vorticity propagation equation
(\ref{a2}), and identity (\ref{a13}). Thus
\be
A_a =\D_a\varphi \,,
\label{3}\ee
where $\varphi$ is the covariant
relativistic generalisation of the Newtonian 
potential. Then the shear constraint becomes
\be
{\cal E}_{ab}\equiv E_{ab} - \D_{\la a}\D_{b\ra}\varphi=0\,.
\label{q1}
\ee

In summary, the only independent new constraint is (\ref{q1}), 
and any conditions that follow from its derivatives.
What is happening here is that the consistent 
evolution of the basic
constraints ${\cal C}^A$ ($A=1,2,\cdots, 5$) is not affected by
introducing a new constraint. It is the new constraint
${\cal E}$ which leads to integrability conditions.
The freedom in the
gravito-electric field is clearly central to the consistency
of the silent models, and conversely, it is the
longitudinal condition (\ref{q1})
on that field which produces integrability conditions
in the quasi-Newtonian subcase.

\subsection{Time evolution}

The time derivative of Eq. (\ref{q1}) follows from the
gravito-electric propagation equation (\ref{a4}) and the
identity
\[
\left\{\D_{\la a}\D_{b\ra}\varphi\right\}^{\rd}=\D_{\la a}\D_{b\ra}
\dot{\varphi}+\left(\dot{\varphi}-2H\right)\D_{\la a} 
\D_{b\ra}\varphi \,,
\]
which is proved using identities (\ref{a14}) and 
(\ref{a15}) and Eq. (\ref{3}). 
It follows that
\be
\dot{\cal E}_{ab} 
= -3H{\cal E}_{ab}
-{\ts{1\over2}}\D_{\la a}{\cal C}^2{}_{b\ra}
-\left(\dot{\varphi}+H\right)\D_{\la a}\D_{b\ra}
\varphi-\D_{\la a}\D_{b\ra}\left(\dot{\varphi}+{\ts{1\over3}}
\Theta\right) \,,
\label{q1.}\ee
on using Eq. (\ref{a6}).
Thus ${\cal E}_{ab}$ evolves consistently if
\be
\D_{\la a}\D_{b\ra}\left(\dot{\varphi}+{\ts{1\over3}}\Theta\right)
+\left(\dot{\varphi}+{\ts{1\over3}}\Theta\right)\D_{\la a}
\D_{b\ra}\varphi =0 \,.
\label{q1..}\ee
This is the first integrability condition in quasi-Newtonian
cosmologies. It represents an extension of the condition
derived in \cite{vee}, since there a particular ansatz is assumed
a priori for $\dot{\varphi}$, i.e.
\[
\dot{\varphi}=\alpha\Theta \,,
\]
where $\alpha$ is a constant parameter. Choosing $\alpha=-{1\over3}$,
i.e.
\be
\dot{\varphi}+{\ts{1\over3}}\Theta=0 \,,
\label{q3}\ee
it is clear that 
the integrability condition
(\ref{q1..}) is reduced to an identity. 

The equation (\ref{q3}) will be adopted here, since it is
covariant, has a clear geometric motivation (as given in 
\cite{vee}), and guarantees consistent evolution of the 
gravito-electric constraint.
Before proceeding with the van Elst-Ellis solution,
it is interesting to ask whether it may be generalized.
The integrability condition (\ref{q1..}) may be rewritten as
\be
\D_{\la a}\D_{b\ra}\left\{{\rm e}^\varphi
\left(\dot{\varphi}+{\ts{1\over3}}\Theta\right)\right\}=0\,.
\label{q1'}\ee
This shows clearly how
the van Elst-Ellis solution may be generalized to
\be
\dot{\varphi}+{\ts{1\over3}}\Theta=\beta{\rm e}^{-\varphi} \,,
\label{q1''}\ee
where $\D_a\beta=0$, i.e. $\beta$ is an arbitrary
background scalar.
There does not appear to be any advantage in adopting the
generalized solution (\ref{q1''}).
It is not clear whether more
general solutions of the condition may be found.

What are the immediate consequences of 
the van Elst-Ellis solution to
the integrability condition?
Firstly, the time evolution of (\ref{q3}) itself leads to
the covariant
modified Poisson equation
\be
\D^2\varphi={\ts{1\over2}}\rho-\left(3\ddot{\varphi}+
\Theta \dot{\varphi}\right) \,,
\label{q4}\ee
after using the Raychaudhuri equation (\ref{a1}).
This equation governs the relativistic
gravitational potential for a given energy density.

Secondly, one can get a crucial evolution equation for the 
4-acceleration \cite{vee}.
Such an evolution equation is not present in
the set of general evolution equations. It arises via
the shearfree condition, as a consequence of Eq. (\ref{q3}).
Taking the gradient of Eq. (\ref{q3}), and using identity
(\ref{a14}) and the div $\sigma$ constraint
(\ref{a6}), one gets\footnote{
Note that the same evolution equation (\ref{q4'}) is obtained if
the generalized solution (\ref{q1''}) is used, with
$\beta\neq 0=\D_a\beta$.}
\be
\dot{A}_a+2HA_a=-{\ts{1\over2}}\rho v_a    \,.
\label{q4'}\ee
Now there is also an evolution equation for $v_a$ \cite{vee}:
\be
\dot{v}_a+Hv_a=-A_a \,,
\label{q17}\ee
as follows from the conservation
equations (\ref{c3}) and (\ref{c4}), or from the comoving frame
equations (\ref{8}) and (\ref{12}).
This is
just the relativistic generalization of the Newtonian equation
for relative acceleration:
\[
{d\vec{v}\over dt}=-\vec{\nabla}\varphi \,.
\]
The coupled evolution equations (\ref{q4'}) and (\ref{q17})
may be decoupled to produce second order equations in either
quantity. For $v_a$:
\be
\ddot{v}_a+3H\dot{v}_a-\left(H^2+{2{\cal K}\over a^2}\right)v_a=0\,,
\label{q18}\ee
on using Eq. (\ref{frw}), while for $A_a$:
\be
\ddot{A}_a+6H\dot{A}_a+{\ts{1\over2}}\left(7H^2-{5{\cal K}
\over a^2}\right)A_a=0 \,.
\label{q18'}\ee
These equations may be solved to find the velocity perturbations
$v_a$ and the 4-acceleration $A_a$.
Since there are no spatial derivatives in Eq. (\ref{q18}), the 
velocity
perturbations are independent of scale.

For a flat background, with ${\cal K}=0$ and $H\propto a^{-3/2}$,
equation (\ref{q18})
is readily solved analytically:
\be
v_a = \Lambda^{(+)}_a\,a^{1/2}+\Lambda^{(-)}_a\,a^{-2}\,,
\label{q19}\ee
where $(+)$ and $(-)$ denote the growing and decaying modes, and
$\dot{\Lambda}^{(\pm)}_a=0$. Using Eq. (\ref{q17}), the solution
for the acceleration follows as
\be
A_a = -\Lambda^{(+)}_a\,a^{-1}+{\ts{2\over3}}
\Lambda^{(-)}_a\,a^{-7/2} \,.
\label{q19'}
\ee

Finally, one can derive another equation for $\varphi$ using
Eq. (\ref{q3}), which implies, together with Eq. (\ref{a6}), that
\be
v_a=-{2\over\rho}\D_a\dot{\varphi} \,.
\label{q12}\ee
Then using Eq. (\ref{q12}) in Eq. (\ref{q18}), it follows that
\be
\D_a\ddot{\varphi}+3H\D_a\dot{\varphi}-{\ts{1\over3}}(\rho-3H^2)
\D_a\varphi=0 \,.
\label{q12'}\ee
Note that the background energy conservation equation and
Eq. (\ref{q12}) determine the velocity potential in terms
of the gravitational potential:
\[
v_a=\D_a\psi ~\mbox{ where }~ \psi=-\left({2\over\rho_0a_0^3}\right)
a^3\dot{\varphi} \,.
\]

\subsection{Spatial consistency}

Spatial derivatives of the new constraint ${\cal E}$
determine whether
it is consistent with ${\cal C}^A$ on an initial spatial
surface.
Taking the curl of Eq. (\ref{q1}) produces
no restrictions, since $\c E_{ab}=0$ by the gravito-magnetic
propagation equation
(\ref{a5}) and constraint equation (\ref{a8}), and since
$\c \D_{\la a}\D_{b\ra}\varphi$ vanishes identically.
This follows from using identities (\ref{a13}) and (\ref{a16}):
\begin{eqnarray*}
\c\D_{\la a}\D_{b\ra}\varphi &=& \c\D_{(a}\D_{b)}\varphi
=\c\D_a\D_b\varphi \\
{}&=& \ep_{cd(a}\D^{[c}\D^{d]}\D_{b)}\varphi\\
{}&=& \left(H^2-{\ts{1\over3}}\rho\right)\ep_{cd(a}h^d{}_{b)}
\D^c\varphi=0 \,.
\end{eqnarray*}

The divergence however gives a nontrivial 
condition. First, one needs an identity for the divergence of the
distortion:
\be
\D^b\D_{\la a}A_{b\ra}={\ts{1\over2}}\D^2A_a+{\ts{1\over6}}\D_a
\left(\div A\right)+{\ts{1\over3}}\left(\rho-3H^2\right)A_a \,,
\label{q12''}\ee
which holds for any projected vector $A_a$, and
follows from identity (\ref{a16}). Using equations
(\ref{3}) and (\ref{a19}), 
it follows that
\be
\D^b\D_{\la a}\D_{b\ra}\varphi={\ts{2\over3}}\D_a\left(\D^2\varphi
\right)+{\ts{2\over3}}\left(\rho-3H^2\right)\D_a\varphi \,.
\label{q12a}\ee
Now use equations (\ref{q12a}), (\ref{a9}) and (\ref{a6}) to get
\[
\left(\div{\cal E}\right)_a={\cal C}^4{}_a-H{\cal C}^2{}_a
+{\ts{1\over3}}\D_a\rho-{\ts{2\over3}}H\D_a\Theta-{\ts{2\over3}}
\D_a\left(\D^2\varphi\right)-{\ts{2\over3}}\left(\rho-3H^2\right)
\D_a\varphi \,.
\]
Thus there is a second integrability condition arising 
from Eq. (\ref{q1}), i.e.
\be
\D_a\rho-2H\D_a\Theta-2
\D_a\left(\D^2\varphi\right)-2\left(\rho-3H^2\right)
\D_a\varphi=0 \,.
\label{new}\ee
In general, this appears to be independent of the
first integrability condition (\ref{q1..}). However, if one uses
the van Elst-Ellis solution (\ref{q3}) of the first condition,
then the second condition is identically satisfied.
This can be seen as follows.
Taking the gradient of Eq. (\ref{q4}) and using Eq.
(\ref{q3}), one finds
that the second integrability condition (\ref{new}) becomes
\be
2\left[\D_a\ddot{\varphi}+3H\D_a\dot{\varphi}
-{\ts{1\over3}}\left(\rho-3H^2\right)\D_a\varphi\right]=0\,,
\label{q12b}\ee
which is identically satisfied by virtue of Eq. (\ref{q12'}).

\section{Density perturbations}

In \cite{vee} the density perturbation
equation is found
in the quasi-Newtonian frame, which requires incorporating
a complicated energy flux source term. 
A simple alternative is to 
work in the comoving frame, which leads directly to
solutions of the density perturbation equation in terms of the
velocity perturbations.

The covariant
density perturbation scalar \cite{eb} in the comoving frame is
\be
\t{\delta}=a\t{\D}^a\t{\delta}_a~\mbox{ where }
~\t{\delta}_a={a\t{\D}_a\rho 
\over\rho} \,.\label{q7'}
\ee
Using Eq. (\ref{q7'}) and the identity
\[
\t{\D}_af = \D_af+\dot{f}v_a \,,
\]
it follows that to linear order the density perturbation scalars
in the comoving and quasi-Newtonian frames are related by
\be
{\delta}=\t{\delta}+3a^2H\D^av_a\,.
\label{q8}\ee

Covariant time derivatives in the comoving 
and quasi-Newtonian frames agree to linear order:
$\t{u}^a\nabla_a S=\dot{S}$. Thus
$\t{\delta}$ satisfies the standard equation 
(with ${\cal K}=0$) for dust \cite{eb}
\be
(\t{\delta})^{\!\rd\rd}+2H(\t{\delta})^{\rd}
-{\ts{3\over2}}H^2\t{\delta}  =0 \,.
\label{q7}\ee
Since the solution $\t{\delta}$ of Eq. (\ref{q7}) is well known, one 
can use Eq. (\ref{q8}) to write down the density perturbations
in the quasi-Newtonian 
frame:
\be
\delta =\Delta^{(+)}\,a+\Delta^{(-)}\,a^{-3/2}
+\gamma a^{-1/2}{\cal V} \,,
\label{q9}\ee
where $\dot{\Delta}^{(\pm)}=0$, and
$\gamma=2a_1^{3/2}$ is a constant, with 
$a_1=a/t^{2/3}$, and
\[
{\cal V}=a\D^av_a
\] 
is the comoving divergence of the peculiar
velocity field. This quantity encodes the scalar contribution
of velocity perturbations to density perturbations.
Using the solution (\ref{q19}) for $v_a$, one finds that 
\be
{\cal V}=\Gamma^{(+)}\,a^{1/2}+\Gamma^{(-)}\,a^{-2}
~\mbox{ where }~\Gamma^{(\pm)}=a\D^a\Lambda^{(\pm)}_a \,.
\label{q10}\ee
It follows from the identity (\ref{a15}) that $\dot{\Gamma}^\pm=0$. 
Then using Eq. (\ref{q10}) in Eq.
(\ref{q9}) gives
\be
\delta =\Delta^{(+)}\,a+\Delta^{(-)}\,a^{-3/2}
+\gamma \Gamma^{(+)} +\gamma\Gamma^{(-)}\,a^{-5/2}\,.
\label{q9'}\ee
The correction to the standard solution affects all scales.
The growing mode is shifted by an amount that is a 
comoving constant, which will be dominated by the term that
grows as $a$.
A similar feature arises in 
a non-covariant Lagrangian perturbation analysis 
(see \cite{tf}).
The decaying mode (not considered in \cite{tf})
is corrected by an amount which dies away
more rapidly, and so becomes negligible.

\section{Concluding remarks}

The covariant approach to quasi-Newtonian models developed
by van Elst and Ellis \cite{vee}
has been applied and extended here, in
order to derive and solve the equations governing density and
velocity perturbations. The quasi-Newtonian
zero-shear hypersurfaces gauge (or longitudinal gauge)
implicitly involves integrability conditions -- i.e. it
incorporates a dynamical condition, and is not pure gauge.
This has been
overlooked in some non-covariant treatments, effectively
amounting to gauge-dependent assumptions about relative-velocity
effects. A fully covariant general relativistic treatment
has uncovered these integrability conditions and their
implication for the perturbations. The perturbations have been
determined, making explicit the relative-velocity effects.

The integrability condition found in \cite{vee} was
generalized, via a simple approach that showed how the
quasi-Newtonian models are in fact a sub-class of the
linearized silent models. The generalized first integrability
condition (\ref{q1'}) naturally leads to the generalization
(\ref{q1''})
of the van Elst-Ellis solution (\ref{q3}).
Furthermore, a second integrability condition (\ref{new})
was found by investigating spatial consistency.
The integrability conditions were shown to
be crucial for determining the perturbations, since the solution
(\ref{q3})
of the integrability conditions implies a propagation equation
(\ref{q4'}) for the 4-acceleration. In turn, this propagation
equation leads to the velocity perturbation equation
(\ref{q18}), and the solution (\ref{q19}) was given for
a flat background. By transforming to the comoving frame,
a simple relation (\ref{q8}) was derived for the density
perturbations, which were given by Eq. (\ref{q9}) 
in terms of the velocity perturbations
and the standard dust solution for a flat background. The
density perturbations were given analytically by Eq. (\ref{q9'}),
which showed how relative-velocity effects produce small corrections
to both the growing and decaying modes on all scales.

These results underline the importance of general relativistic
constraint equations and the integrability conditions that
can arise from imposing various physical and geometric 
assumptions in cosmology. A covariant approach \cite{e,eb} avoids
many of the intricate problems arising from gauge-dependent
approaches.
The improved
covariant formalism of \cite{m}, summarized in section II
and supplemented by the identities of appendix A and the new
results in appendix B on transformations of the covariant
quantities, has been central to deriving these results.
Similar methods can in principle be applied to investigate
the consistency and underlying implications of other
gauge choices or of various approximations (such as the Zel'dovich
approximation)
in general relativistic cosmological perturbations. This is a
subject of further research.

\newpage

\appendix

\section{Linearized identities}

Useful differential identities \cite{mt}:
\begin{eqnarray}
\c\D_af &=&-2\dot{f}\omega_a \,,
\label{a13} \\
\D^2\left(\D_af\right) &=&\D_a\left(\D^2f\right) 
+{\ts{2\over3}}\left(\rho-3H^2\right)\D_a f+2\dot{f}\c\omega_a
\,, \label{a19}\\
\left(\D_af\right)^{\rd} &=& \D_a\dot{f}-H\D_af+\dot{f}A_a \,,
\label{a14}\\
\left(\D_aS_{b\cdots}\right)^{\rd} &=& \D_a\dot{S}_{b\cdots}
-H\D_aS_{b\cdots}\,,
\label{a15}\\
\left(\D^2 f\right)^{\rd} &=& \D^2\dot{f}-2H\D^2 f 
+\dot{f}\div A \,,\label{a21}\\
\D_{[a}\D_{b]}V_c &=& 
{\ts{1\over3}}\left(3H^2-\rho\right)V_{[a}h_{b]c}
\,, \label{a16}\\
\D_{[a}\D_{b]}S^{cd} &=& {\ts{2\over3}}
\left(3H^2-\rho\right)
S_{[a}{}^{(c}h_{b]}{}^{d)} \,, \label{a17}\\
\div\c V &=& 0 \label{a20}\\
(\div\c S)_{a} &=& {\ts{1\over2}}\c (\div S)_{a}\,,
\label{a18}\\
\c\c V_a &=& \D_a (\div V)
-\D^2V_a+{\ts{2\over3}}\left(\rho-3H^2\right)V_a \,,
\label{a22}\\
\c\c S_{ab} &=& {\ts{3\over2}}\D_{\la a}(\div S)_{b\ra}-\D^2S_{ab}
+\left(\rho-3H^2\right)S_{ab} \,,
\label{a23}
\end{eqnarray}
where the vectors and tensors vanish in the background,
$S_{ab}=S_{\la ab\ra}$, and all identities
except (\ref{a13}) are linearized. (Nonlinear identities are given
in \cite{m,ve,mes}.)

\section{Transformations under change of frame}

Change in 4-velocity:
\be
\tilde{u}_a = \gamma(u_a+v_a) ~~\mbox{ where }~~
v_au^a=0\,,~\gamma=(1-v^2)^{-1/2} \,.
\label{b1}\ee
The following
algebraic relations are needed:
\begin{eqnarray*}
g_{ab} &=& h_{ab}-u_au_b
= \t{h}_{ab}-\t{u}_a\t{u}_b \,,\\
\t{h}_{ab} &=& h_{ab}+\gamma^2\left[v^2u_a u_b+2u_{(a}v_{b)}
+v_a v_b\right]\,, \\
\eta_{abcd} &=& 2u_{[a}\ep_{b]cd}-2\ep_{ab[c}u_{d]}
= 2\tilde{u}_{[a}\tilde{\ep}_{b]cd}-
2\tilde{\ep}_{ab[c}\tilde{u}_{d]} \,,\\
\tilde{\ep}_{abc} &=& \gamma\ep_{abc}+\gamma\left\{2u_{[a}\ep_{b]cd} 
+u_c\ep_{abd}\right\}v^d \,, \\
C_{ab}{}{}^{cd} &=& 
4\left\{u_{[a}u^{[c}+h_{[a}{}^{[c}\right\}E_{b]}{}^{d]}
+2\ep_{abe}u^{[c}H^{d]e}+2u_{[a}H_{b]e}\ep^{cde} \\
&=&
4\left\{\tilde{u}_{[a}\tilde{u}^{[c}+\tilde{h}_{[a}{}^{[c}\right\}
\tilde{E}_{b]}{}^{d]}
+2\tilde{\ep}_{abe}\tilde{u}^{[c}\tilde{H}^{d]e}+2
\tilde{u}_{[a}\tilde{H}_{b]e}\tilde{\ep}^{cde} \,,
\end{eqnarray*}
together with the decomposition \cite{mes}
\begin{eqnarray}
\nabla_bv_a &=& -u_b\left\{\dot{v}_{\la a\ra}+A_cv^cu_a\right\}+
u_a\left\{{\ts{1\over3}}\Theta v_b+\sigma_{bc}v^c+[\omega,v]_b\right\}
\nonumber\\
{}&&{}+{\ts{1\over3}}\left(\div v\right)h_{ab}-{\ts{1\over2}}
\ep_{abc}\c v^c+\D_{\la a}v_{b\ra} \,,
\label{b2}\end{eqnarray}
where $[W,V]_a\equiv
\ep_{abc}W^bV^c$. Then the following exact nonlinear transformations
may be derived.

Kinematic quantities
(using
$\nabla_a\gamma=\gamma^3v^b\nabla_av_b$):
\bea
\t{\Theta} &=& \gamma\Theta+\gamma\left(\div v+A^av_a\right)
+\gamma^3W \,,\label{b3}\\
\t{A}_a &=& \gamma^2A_a+\gamma^2\left\{
\dot{v}_{\la a\ra}+
{\ts{1\over3}}\Theta v_a+
\sigma_{ab}v^b-[\omega,v]_a
+\left({\ts{1\over3}}\Theta v^2+A^bv_b+\sigma_{bc}v^bv^c
\right)u_a
\right.\nonumber\\
&&\left. 
+{\ts{1\over3}}(\div v)v_a+{\ts{1\over2}}[v,\c v]_a
+v^b\D_{\la b}v_{a\ra}\right\} 
+\gamma^4W(u_a+v_a) \,, \label{b4}
\\
\t{\omega}_a &=& \gamma^2\left\{\left(1-{\ts{1\over2}}v^2\right)
\omega_a- {\ts{1\over2}}\c v_a+{\ts{1\over2}}v_b\left(2\omega^b-\c v^b
\right)u_a+{\ts{1\over2}}v_b\omega^bv_a \right.
\nonumber\\
&&\left.{}+{\ts{1\over2}}[A,v]_a+{\ts{1\over2}}[\dot{v},v]_a 
+{\ts{1\over2}}\ep_{abc}\sigma^b{}_dv^cv^d \right\}
\,, \label{b5} \\
\t{\sigma}_{ab} &=& \gamma\sigma_{ab}
+\gamma(1+\gamma^2)u_{(a}\sigma_{b)c}v^c
+\gamma^2A_{(a}\left[v_{b)}+v^2u_{b)}\right] \nonumber\\
&&{}+ \gamma\D_{\la a}v_{b\ra} 
-{\ts{1\over3}}h_{ab}\left[
A_cv^c+\gamma^2\left(W-\dot{v}_cv^c\right)\right]
\nonumber\\
&&{}+\gamma^3u_au_b\left[\sigma_{cd}v^cv^d+{\ts{2\over3}}v^2A_cv^c
-v^cv^d\D_{\la c}v_{d\ra}+\left(\gamma^4-{\ts{1\over3}}v^2\gamma^2
-1\right)W\right]\nonumber\\
&&{}+\gamma^3u_{(a}v_{b)}\left[A_cv^c+\sigma_{cd}v^cv^d-\dot{v}_cv^c+
2\gamma^2\left(\gamma^2-{\ts{1\over3}}\right)W\right] \nonumber\\
&&{}+{\ts{1\over3}}\gamma^3v_av_b\left[\div v-A_cv^c
+\gamma^2\left(3\gamma^2-1\right)W\right]+\gamma^3v_{\la a}
\dot{v}_{b\ra}+v^2\gamma^3u_{(a}\dot{v}_{\la b\ra)} \nonumber\\
&&{}+\gamma^3v_{(a}\sigma_{b)c}v^c-\gamma^3[\omega,v]_{(a}
\left\{v_{b)}+v^2u_{b)}\right\}
+2\gamma^3v^c\D_{\la c}v_{(a\ra}\left\{v_{b)}+u_{b)}\right\}
\,, \label{b6}
\eear
where
$W\equiv
\dot{v}_cv^c+{\ts{1\over3}}v^2\div v+v^cv^d\D_{\la c}v_{d\ra}$.
         
Dynamic quantities (compare \cite{mw}):
\bea
\t{\rho} &=& \rho+\gamma^2\left[v^2(\rho+p)-2q_av^a+\pi_{ab}v^av^b
\right]\,, \label{b7}\\
\t{p} &=& p+{\ts{1\over3}}\gamma^2\left[v^2(\rho+p)-2q_av^a
+\pi_{ab}v^av^b\right]\,, \label{b8}\\
\t{q}_a &=& \gamma q_a-\gamma\pi_{ab}v^b-\gamma^3\left[(\rho+p)
-2q_bv^b+\pi_{bc}v^bv^c\right]v_a \nonumber\\
{}&&-\gamma^3\left[v^2(\rho+p)-(1+v^2)q_bv^b+\pi_{bc}v^bv^c\right]u_a
\,, \label{b9} \\
\t{\pi}_{ab} &=& \pi_{ab}+2\gamma^2v^c\pi_{c(a}\left\{u_{b)}+
v_{b)}\right\}-2v^2\gamma^2q_{(a}u_{b)}-2\gamma^2q_{(a}v_{b)}
\nonumber\\
{}&&-{\ts{1\over3}}\gamma^2\left[v^2(\rho+p)-2q_cv^c+\pi_{cd}v^cv^d
\right]h_{ab} \nonumber\\
&&{}+{\ts{1\over3}}\gamma^4\left[2v^4(\rho+p)-4v^2q_cv^c
+(3-v^2)\pi_{cd}v^cv^d\right]u_au_b \nonumber\\
{}&&+{\ts{2\over3}}\gamma^4\left[2v^2(\rho+p)-(1+3v^2)q_cv^c+2
\pi_{cd}v^cv^d\right]u_{(a}v_{b)} \nonumber\\
&&{}+{\ts{1\over3}}\gamma^4\left[(3-v^2)(\rho+p)-4q_cv^c+2
\pi_{cd}v^cv^d\right]v_av_b \,. \label{b10}
\eear

Gravito-electromagnetic field:
\begin{eqnarray}
\tilde{E}_{ab} &=&
\gamma^2\left\{(1+v^2)
E_{ab}+v^c\left[2\ep_{cd(a}H_{b)}{}^d+
2E_{c(a}u_{b)}\right.\right.\nonumber\\
&&\left.\left.{}+(u_au_b+h_{ab})E_{cd}v^d-2E_{c(a}v_{b)}
+2u_{(a}\ep_{b)cd}H^{de}v_e\right]\right\}\,, \label{b11}\\
\tilde{H}_{ab} &=&
\gamma^2\left\{(1+v^2)
H_{ab}+v^c\left[-2\ep_{cd(a}E_{b)}{}^d+
2H_{c(a}u_{b)}\right.\right.\nonumber\\
&&\left.\left.{}+(u_au_b+h_{ab})H_{cd}v^d-2H_{c(a}v_{b)}
-2u_{(a}\ep_{b)cd}E^{de}v_e\right]\right\} \,. \label{b12}
\end{eqnarray}

\end{document}